\documentclass[conference,10pt]{IEEEtran}
\IEEEoverridecommandlockouts

\usepackage{ulem}
\usepackage{url}
\usepackage{float}
\usepackage{stfloats}

\usepackage{cite}

\usepackage{graphicx}
\usepackage{amsmath, amsthm}
\usepackage{amssymb}
\usepackage{mathtools}
\usepackage{comment}
\usepackage[subrefformat=parens,labelformat=parens]{subfig}
\usepackage{bm}
\usepackage{multirow}
\usepackage{threeparttable,booktabs}
\usepackage{tikz}
\usepackage[mathcal]{eucal}
\usepackage[ruled,lined,linesnumbered,noend]{algorithm2e}
\usepackage{filecontents}                                  
\usepackage{pgfplots}
\usepackage{pgfplotstable}
\pgfplotsset{compat=newest}

\theoremstyle{plain}

\newtheorem*{theorem*}{\textbf{Theorem}}

\theoremstyle{definition}

\newtheorem*{problem*}{\textbf{Problem}}

\usepackage[skip=1pt]{caption}            
\graphicspath{{./figs/}}

\begin{document}

\title{
\textbf{
  HeteroSTA: A CPU-GPU Heterogeneous Static Timing Analysis Engine with Holistic Industrial Design Support
}}

\iftrue
\author{
  \IEEEauthorblockN{Zizheng Guo\textsuperscript{1,2}, Haichuan Liu\textsuperscript{1}, Xizhe Shi\textsuperscript{1}, Shenglu Hua\textsuperscript{1}, Zuodong Zhang\textsuperscript{2}, Chunyuan Zhao\textsuperscript{1}, \\Runsheng Wang\textsuperscript{1,2,3}, Yibo Lin\textsuperscript{1,2,3,*}}
  \IEEEauthorblockA{
    \textsuperscript{1}School of Integrated Circuits, \textit{Peking University}, \textsuperscript{2}Institute of EDA, \textit{Peking University}\\
    \textsuperscript{3}Beijing Advanced Innovation Center for Integrated Circuits, *Corresponding author: yibolin@pku.edu.cn
  }
}
\fi



\maketitle

\bstctlcite{ieee:norepeatednames, ieee:etal8}

\begin{abstract}
  We introduce in this paper, HeteroSTA, the first CPU-GPU heterogeneous timing analysis engine that efficiently supports: (1) a set of delay calculation models providing versatile accuracy-speed choices without relying on an external golden tool, (2) robust support for industry formats, including especially the .sdc constraints containing all common timing exceptions, clock domains, and case analysis modes, and (3) end-to-end GPU-acceleration for both graph-based and path-based timing queries, all exposed as a zero-overhead flattened heterogeneous application programming interface (API). HeteroSTA is publicly available with both a standalone binary executable and an embeddable shared library targeting ubiquitous academic and industry applications. Example use cases as a standalone tool, a timing-driven DREAMPlace 4.0 integration, and a timing-driven global routing integration have all demonstrated remarkable runtime speed-up and comparable quality.
\end{abstract}


\begin{figure*}[bp]
  \setlength{\linewidth}{\textwidth}
  \setlength{\hsize}{\textwidth}
  \centering
  \includegraphics[width=\linewidth]{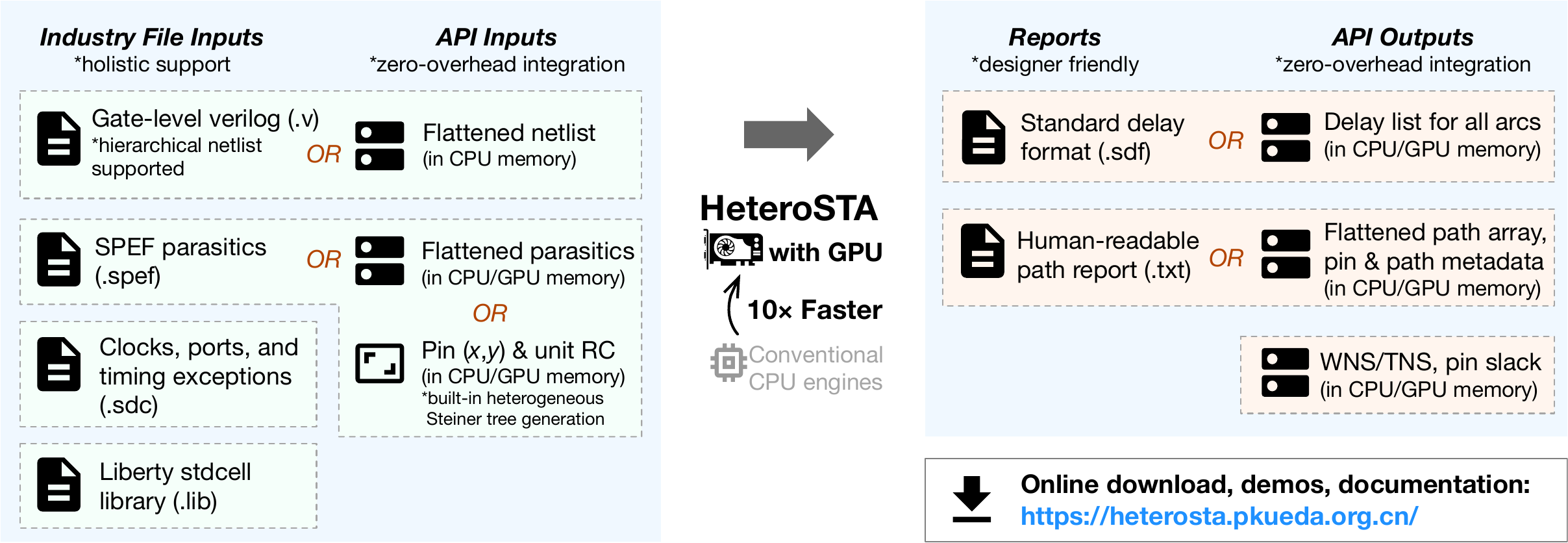}
  \caption{
    HeteroSTA supports various input and output combinations featuring holistic industry file format support and a zero-cost abstraction interface.
    It delivers an order of magnitude STA speed-up for heterogeneous EDA applications.
  }
  \label{fig:inputs-outputs}
\end{figure*}

\section{Introduction}
\label{sec:Introduction}


With the sheer increase in the VLSI design volume as well as the growing demand for fast turnaround time and rapid design iteration,
heterogeneous CPU/GPU-accelerated EDA has garnered wide attention as an extremely important aspect of next-generation EDA systems.
With years of development, GPU acceleration techniques have covered most of the major EDA stages,
including RTL simulation~\cite{qian2011gpurtl, lin2022rtlflow, zhang2024gl0am, gemdac25},
logic synthesis~\cite{novelrewrite2022, iccadcadcontest2021c_logicrewrite, liu2023rethinking, sun2024massively, finemap2024, groot2025},
partitioning~\cite{goodarzi2016gpupart, liang2024medpart, lee2024gkway, wu2025ghypart, lee2025hyperg, lee2025igkway},
placement~\cite{DREAMPlace, ABCDPlace, guo2022difftdp, agnesina2023autodmp, liu2023xplace, xplace_tcad, kahng2024dg, fusionsizericcad24, fu2024xplace_t, dpo2025, sroadpgpu2025},
routing~\cite{lin2022gamer, gpufluteiccad22, liu2022fastgr, lin2024instantgr, xiao2025instantgr, li2024dgr, zhao2024helemgr, liang2025ispd, zhao2025gta},
design rule checking~\cite{he2022xcheck, he2023opendrc},
timing signoff~\cite{guo20gpusta, arnoldidate24, lin2024gcstimer, gputimertcad23, instadac25, incregpusta2025, heteroexcepticcad24, guo21heterocppr, guo21gpupba, gpupbatcad23, pathgenaspdac25}, etc.
Among these efforts, static timing analysis (STA) is one of the central tasks because all major stages embed STA in their timing optimization inner loop, where STA is invoked thousands of times.

The advancement in heterogeneous CPU/GPU STA algorithms has brought orders-of-magnitude speedup to timing analysis,
covering all major steps in a typical and complete STA engine (delay calculation~\cite{guo20gpusta, arnoldidate24, lin2024gcstimer}, graph propagation~\cite{gputimertcad23, instadac25, incregpusta2025}, exception handling~\cite{heteroexcepticcad24}, and path search~\cite{guo21heterocppr, guo21gpupba, gpupbatcad23, pathgenaspdac25}).
Since then, we have witnessed the early-stage development of timing-driven EDA flows that are end-to-end heterogeneous~\cite{guo2022difftdp, fusionsizericcad24, fu2024xplace_t, dpo2025}.
However, existing research all focuses on the acceleration of specific STA steps without integrating all algorithms into a rich-featured, readily-available heterogeneous STA library.
Prior timing-driven EDA flows were thus all forced to implement their own small proof-of-concept heterogeneous STA modules that have poor design compatibility (e.g., industrial Verilog, SDC, and SPEF), low analysis accuracy, and inferior runtime efficiency.

The gap between research and application in the GPU-accelerated STA field has motivated us to develop a holistic heterogeneous STA library, \textbf{HeteroSTA}, targeting widespread academic and industry adoption. With such a library available to the public, both researchers and engineers can bootstrap their research and development of EDA algorithms and flows that benefit greatly from the much faster encapsulated STA function, without having to deal with the enormous engineering details in implementing a full STA engine.

Compared to STA libraries available prior to this work, e.g., OpenTimer~\cite{huang2020opentimerv2}, OpenSTA~\cite{OpenSTA}, GCS-Timer~\cite{lin2024gcstimer}, and INSTA~\cite{instadac25}, HeteroSTA is the first heterogeneous STA engine with holistic industrial design support. It is \emph{self-contained}: it does not require an external golden tool to perform delay calculation or exception preprocessing. It is \emph{industry-compatible}: it accepts industry-standard file formats like gate-level Verilog (with hierarchical support), SPEF parasitics, SDC constraints (covering clocks, ports, and all common timing exception definitions), and the Liberty cell library. It is \emph{integration-ready}: as a library and a set of C/C++ API (Figure~\ref{fig:inputs-outputs}), it is ready to be embedded into different applications, with a novel heterogeneous API design ensuring zero-overhead in data communication between tools and the STA engine.

HeteroSTA is made available from today and can be downloaded at this link: \footnote{https://heterosta.pkueda.org.cn/}, with online documentation and discussion available. Together with the release of the library, we also release two demos integrating HeteroSTA into the open-source timing-driven DREAMPlace 4.0~\cite{liao2023dreamplace} and Efficient-TDP~\cite{shi2025timing} placers, both available on GitHub\footnote{https://github.com/limbo018/DREAMPlace, and https://github.com/PKU-IDEA/Efficient-TDP-HeteroSTA}. We also integrated HeteroSTA into a GPU-accelerated global routing flow. After replacing the original timers in these representative flows with HeteroSTA, we achieve significant end-to-end speedup without quality degradation.


The rest of this paper is organized as follows.
Section~\ref{sec:Preliminary} discusses the challenges of a heterogeneous STA library with a review of current STA tools available.
Section~\ref{sec:Algorithm} demonstrates design details of HeteroSTA.
Section~\ref{sec:Results} presents the experimental results including the runtime performance and accuracy of HeteroSTA as both a standalone timer and an integrated STA engine in timing-driven placement and global routing workloads.
Finally, Section~\ref{sec:Conclusion} concludes the paper.

%


\section{Challenges of a Heterogeneous STA Library}
\label{sec:Preliminary}

\begin{figure}[t]
  \centering
  \includegraphics[width=\linewidth]{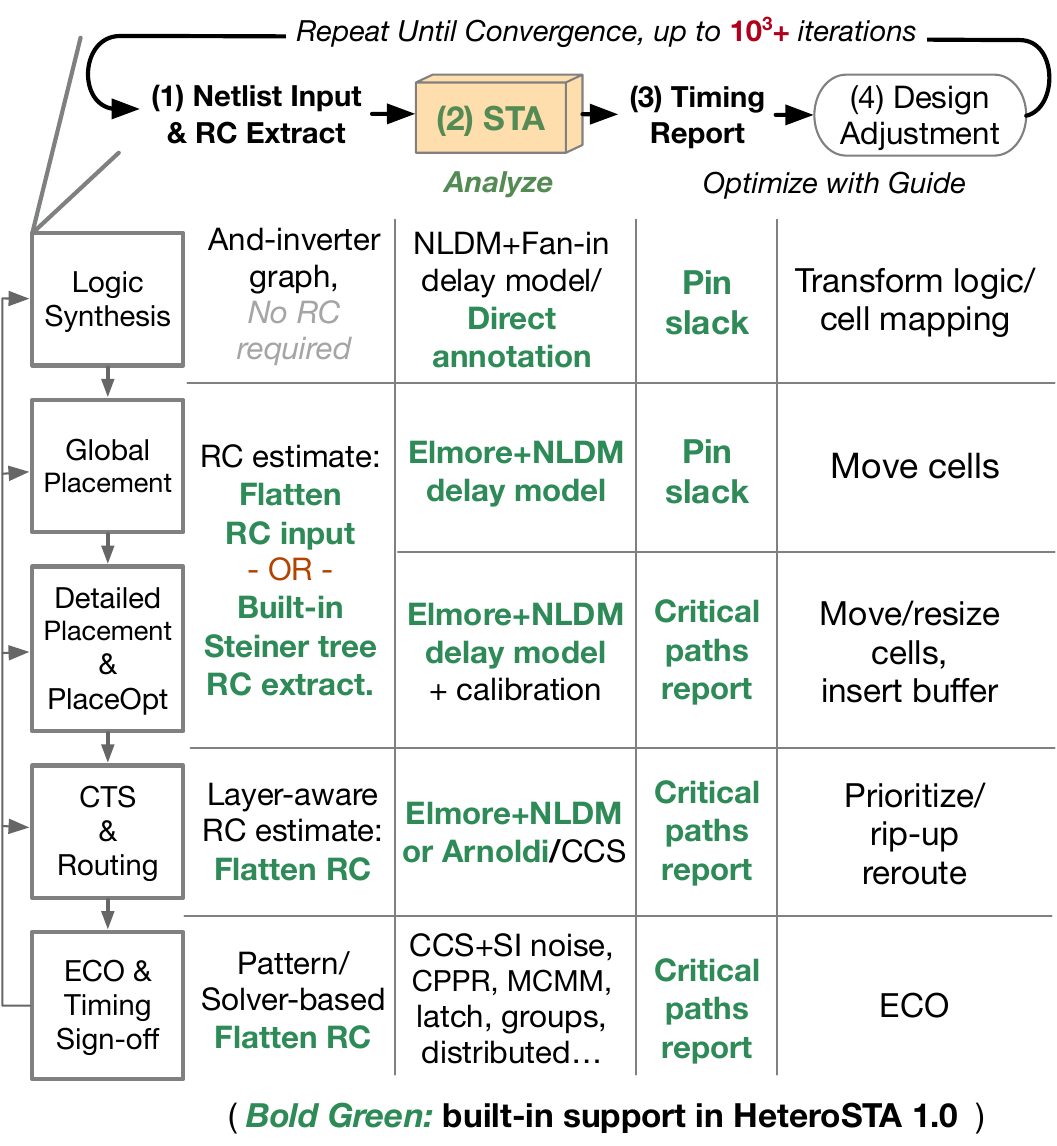}
  \caption{STA is heavily used in design flows: a fast timer is critical in speeding up design iterations. Different stages may require customized STA inputs, outputs, and functions.}
  \label{fig:sta-in-flow}
\end{figure}

STA determines circuit performance by providing graph-based and path-based timing criticality reports given a circuit netlist, a cell library, a list of clocks and exceptions, and a parasitics annotation~\cite{STABook}.
In a typical EDA flow, STA is itself an essential stage managing timing signoff post physical design.
More importantly, STA is integrated as a subprocess in many other stages to guide circuit performance optimization -- such optimizations are usually formulated as an ``analyze-then-optimize'' loop, where STA plays a central role in locating timing bottlenecks (Figure~\ref{fig:sta-in-flow}).

Due to their frequent reuse, core STA engines are packaged as libraries that can be embedded into other flows via an application programming interface (API).
Such APIs are often kept for internal use in commercial EDA companies, and their STA engines (e.g., PrimeTime~\cite{TOOL_pt} and Tempus~\cite{TOOL_tempus}) are only released as standalone tool executables with which interactions are mostly text-based (files or Tcl scripts).
Instead, academic EDA flows often embed open-source STA libraries like OpenSTA~\cite{OpenSTA} and OpenTimer~\cite{huang2020opentimerv2} instead when commercial tools are unavailable and to avoid inefficiencies in text-based interfaces.

\uline{The timing reports mismatch between the golden commercial STA engines and the currently available, embeddable STA libraries} is one of the major challenges that HeteroSTA tries to resolve.
We identify two core issues that are accountable for the mismatch: \emph{delay model accuracy} and \emph{timing exception compatibility}.
Accurately compute the cell and net delays at sub-micron nodes requires advanced circuit models beyond the current widely-used Elmore delay model.
Timing exceptions such as false paths, multi-cycle paths, case analysis, and cross clock region paths (usually defined in \texttt{.sdc} files) can significantly complicate the data structures and states in timing propagation implementation and open-source STA engines often have limited or buggy support.

Currently, none of the publicly-available STA libraries have resolved both issues.
The recent work INSTA~\cite{instadac25} has presented remarkable correlation with commercial tools.
However, we note that it does \emph{not} intend to address these issues by itself -- instead, INSTA relies on a golden commercial STA engine to provide delay annotations and then performs timing propagation and slack calculation based on these annotations.
These calculations are only part of the late steps of timing analysis that are closer to the generation of optimization guides.
Furthermore, in our experiments, we will show incorrect results from INSTA when handling practical timing exceptions in timing propagation.



Heterogeneous CPU/GPU-accelerated EDA flows have demonstrated orders-of-magnitude runtime speedup and unprecedented scalability, enabling rapid design iteration, faster time-to-market, and more efficient design space exploration.
Therefore, they have been regarded as promising directions for future EDA systems.
A heterogeneous EDA flow requires a heterogeneous STA library to achieve its peak performance.
STA is itself known as a runtime bottleneck in the optimization loop.
Moreover, without a GPU-accelerated STA engine, the data interface between the GPU-accelerated optimization loop and the CPU-based STA engine would incur large back-and-forth CPU-GPU data movement cost, as we observe in such flows in reality~\cite{liao2023dreamplace, shi2025timing}.

\uline{The demand on a STA library that is \emph{natively heterogeneous} introduces new major challenges on top of existing ones.}
STA consists of many distinct graph-based algorithms that are hard to parallelize with GPU's single-instruction-multiple-threads (SIMT) model. Fortunately, these years have witnessed novel algorithms and data structures that help in bridging the compute architecture gap and making STA massive parallel, but how to wrap these individual contributions into a \emph{comprehensive STA software} remains uncertain.
In addition to the speed of STA itself, the overall efficiency of the heterogeneous-STA-powered EDA flow is heavily determined by the API design of the STA library.
OpenTimer and OpenSTA expose object-oriented programming (OOP) APIs that work on individual cells, nets, or pins.
Such OOP-based APIs are no longer feasible for heterogeneous data communications due to the high translation overhead of data structures~\cite{guo20gpusta}.
Ideally, the communication between heterogeneous STA libraries and EDA flows should be based on a set of \emph{zero-overhead heterogeneous APIs}, which is highly nontrivial without existing practices to follow.



\section{HeteroSTA}
\label{sec:Algorithm}
\begin{figure}[t]
  \centering
  \includegraphics[width=\linewidth]{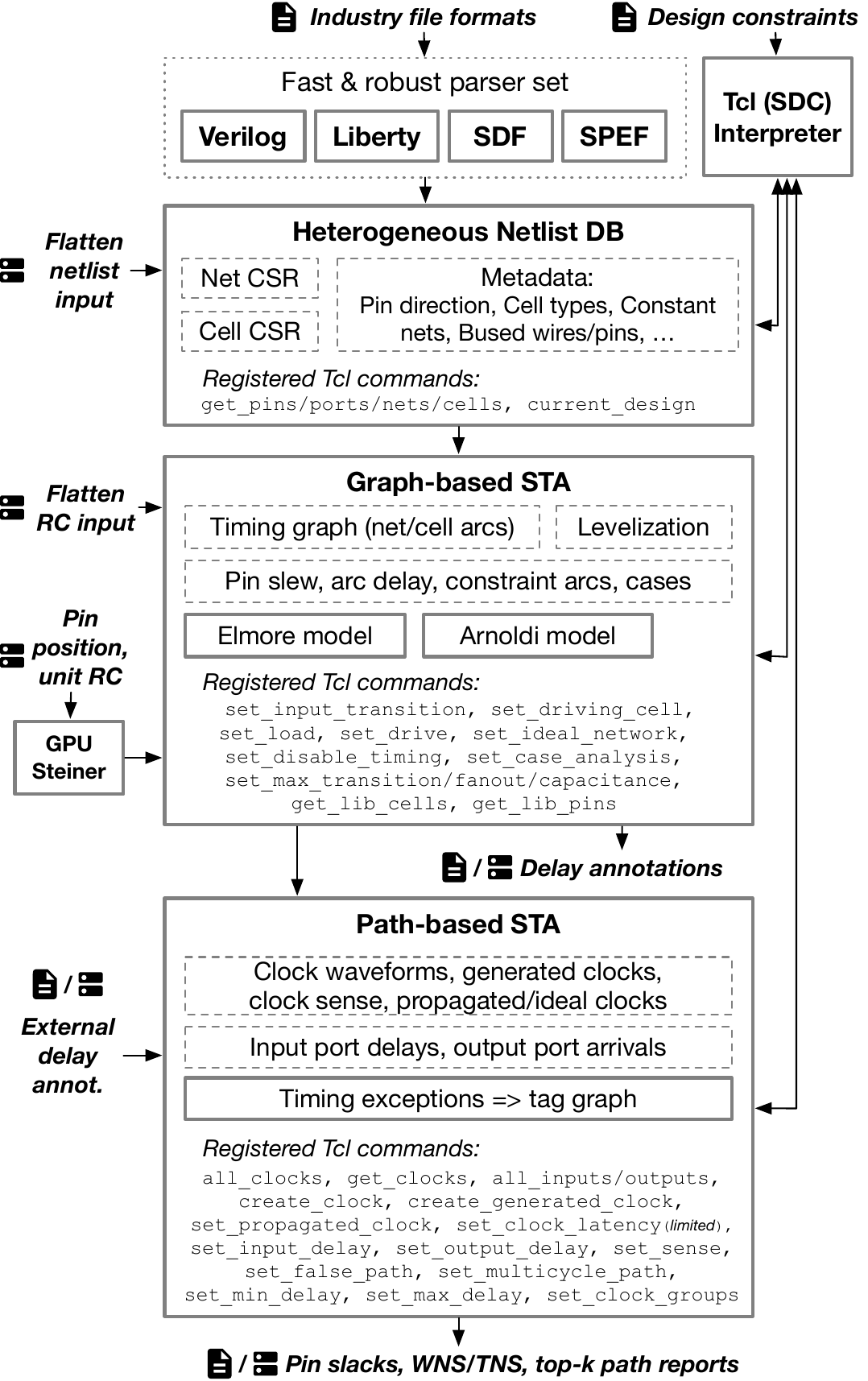}
  \caption{The modular design of HeteroSTA showing inputs, outputs, and data interactions between submodules.}
  \label{fig:arch}
\end{figure}

As presented in Figure~\ref{fig:inputs-outputs}, HeteroSTA features both a versatile support for industrial file formats and a zero-cost data API designed for heterogeneous integration.
To achieve this, HeteroSTA applies a modularized software design consisting of parsers, netlist database, graph-based STA, and path-based STA.
Design constraints are parsed with a built-in Tcl interpreter and dispatched to different levels of abstraction.
Figure~\ref{fig:arch} presents a detailed view of the architecture of HeteroSTA.

\emph{Industry file format parsers.}
We implement a set of high-performance parsers for industry-standard formats, including gate-level Verilog (\texttt{.v}), Liberty cell library (\texttt{.lib}), delay annotation (\texttt{.sdf}), and standard parasitics (\texttt{.spef}).
Our parsers are implemented with byte-level parsing expression grammar (PEG) that features a memory-efficient single-pass parsing strategy.
We support streamed parsing and the parsing of gzipped inputs.
Our gate-level Verilog parser supports hierarchical designs, buses, assignments, and constant wires (e.g., \texttt{1'b0}).
Our Liberty parser is verified on several industrial 7--14 nm PDKs with slew derates, \emph{when}-guarded timing arcs, pin functions, etc.
All parsers support multi-threaded parsing: the Verilog, SDF, and SPEF parsers can run multi-threaded on a single file input by splitting the file heuristically into chunks, while the Liberty parser can read multiple \texttt{.lib} files from modern PDKs in parallel.

\emph{Tcl interpreter-based timing exceptions manager.}
We embed a standalone Tcl script interpreter in HeteroSTA to handle design constraints in \texttt{.sdc} files.
This enables HeteroSTA to process advanced scripted SDCs, such as the ones with expressions, for-loops, condition branches, etc.
SDC commands in the Tcl interpreter are registered by individual HeteroSTA function modules across levels of abstraction (Figure~\ref{fig:arch}).

\emph{Heterogeneous netlist database.}
We store the structure of the netlist as a central gate-level flattened netlist database.
Rather than based on OOP styles, this database is natively flattened represented by compressed sparse rows (CSR) edge lists.
We provide two ways to initialize this database.
By inputting a Verilog, HeteroSTA builds the netlist database itself recursively through the hierarchy.
Alternatively, HeteroSTA accepts external net and cell CSR arrays directly from a heterogeneous EDA flow outside, e.g., DREAMPlace.
The latter way eliminates the need to rebuild the netlist or maintain pin indices mappings for STA invocated by other flows.

\emph{Parasitics (RC) inputs.}
HeteroSTA accepts 3 ways to input parasitics (RC) annotations.
Annotations can be given through industry-standard \texttt{.spef} files with built-in parsers.
For a heterogeneous optimization loop, parasitics are usually generated on-the-fly with an algorithm, so we also provide ways to directly send RC to HeteroSTA without file-based communications.
One general way to do this is through a flattened RC interface, where a user packs the generated RC into a set of predefined CSR structures on either CPU or GPU -- this is useful in timing-driven routing where the flow has control over the layer-based RC generation algorithms.
For the placement context, we also embed a built-in Steiner tree-based RC extraction module based on GPU-accelerated FLUTE~\cite{gpufluteiccad22}.
A placer only needs to provide HeteroSTA with pin positions and unit resistance/capacitance values along x/y directions and HeteroSTA will run built-in FLUTE to generate RC estimations and incorporate them in delay calculation.

\emph{Delay models.}
HeteroSTA currently supports two delay calculators: the simple classical Elmore delay calculator, and an Arnoldi-based reduced-order model.
The Elmore model is the fastest yet not accurate enough in late design stages, where Arnoldi might be a better option.
Our delay calculator framework is extensible and in the future we plan to add CCS delay models in the next release of HeteroSTA.

\emph{Report options.}
HeteroSTA supports a variety of timing reports to fit different design needs, including delay annotations, WNS/TNS, pin slacks, and top-$k$ path reports.
Delay annotations can be in the standard \texttt{.sdf} file output as well as in-memory delay arrays directly.
Path reports can be controlled with top-$k$, per-endpoint report limit, and slack-less-than thresholds.
The format of path report can be human-readable plain texts as well as a flattened CSR-based path pin arrays with slacks and other useful metadata.
All output arrays can be on CPU or GPU at user's option, enabling fully heterogeneous optimization loops where the majority of design data never leaves GPU memory.

%



\section{Experimental Results}
\label{sec:Results}
We release HeteroSTA as a Linux shared library (\texttt{.so}), a set of C header files (\texttt{.h}), example sources, and an online documentation.\footnote{https://heterosta.pkueda.org.cn/documentation}
Under the hood, HeteroSTA is implemented in Rust, C++, and CUDA.
HeteroSTA supports both OpenMP-based CPU multithreading and CUDA-based GPU massive parallelism.
Our experiments aim to evaluate HeteroSTA in two aspects: (1) its correlation with commercial tool PrimeTime, in both \emph{delay calculation accuracy} (Section~\ref{sec:result-dcalc}) and \emph{timing exception correctness} (Section~\ref{sec:result-exception}); (2) its end-to-end performance benefit when integrated into representative heterogeneous GPU-accelerated EDA flows, including two \emph{timing-driven global placement} flows (Section~\ref{sec:result-gp}) and one \emph{timing-driven global routing} flow (Section~\ref{sec:result-gr}).
All our experiments are run on a Linux machine with 64 Intel Xeon Platinum 8358 CPU cores, 1 TB memory, and 8 NVIDIA A100-SXM4-80GB GPUs.
Unless otherwise noted, we run all benchmarks 3 times and report average runtime, using 16 CPU cores and 1 GPU (if applicable), which is when their performance generally saturates.

\subsection{Delay Calculation Accuracy}
\label{sec:result-dcalc}

We test the delay modeling accuracy by comparing the delay annotations generated by HeteroSTA, PrimeTime, and OpenSTA on the TAU 2015 contest benchmarks~\cite{hu2015tau} transpiled to a 14~nm process~\cite{arnoldidate24}.
Both HeteroSTA and OpenSTA use their respective NLDM+Arnoldi delay model, while PrimeTime uses its basic (non-CCS) calculation mode. \footnote{CCS support is in our roadmap for next-version release and is not yet available in HeteroSTA 1.0.}
Table~\ref{tab:dcalc} shows a detailed comparison of both runtime and accuracy.
HeteroSTA achieves better correlation ($R^2$ and MAE) than OpenSTA compared to PrimeTime. Specifically, HeteroSTA achieves an average $R^2$ score of 0.985 and a MAE of 2.34 ps. Our accuracy is also stable and consistent across all designs we have tested.

Thanks to heterogeneous CPU/GPU acceleration, the runtime of HeteroSTA significantly outperformed both PrimeTime and OpenSTA among all benchmarks. We are on average 4.89$\times$ faster than PrimeTime and 10.04$\times$ faster than OpenSTA. On the largest design \texttt{leon2}, we are 7.56$\times$ and 14.53$\times$ faster, respectively.

\begin{table*}[t]
\caption{Comparison of runtime, timing arc delay correlation ($R^2$ and MAE) between PrimeTime, OpenSTA, and HeteroSTA. PrimeTime is set as the golden result. Runtimes are in milliseconds.}
\label{tab:dcalc}
\resizebox{\linewidth}{!}{%
\begin{tabular}{l|rrr|rrrr|rrrr|rrrr}
\toprule
\multirow{2}[2]{*}{Benchmark} & \multicolumn{3}{c|}{Statistics} & \multicolumn{4}{c|}{PrimeTime (16C)} & \multicolumn{4}{c|}{OpenSTA (16C)} & \multicolumn{4}{c}{HeteroSTA (16C + 1 GPU)} \\
      & \multicolumn{1}{c}{\#Cells} & \multicolumn{1}{c}{\#Nets} & \multicolumn{1}{c|}{\#Pins} & \multicolumn{1}{c}{Runtime} & \multicolumn{1}{c}{RTR} & \multicolumn{1}{c}{MAE} & \multicolumn{1}{c|}{R2} & \multicolumn{1}{c}{Runtime} & \multicolumn{1}{c}{RTR} & \multicolumn{1}{c}{MAE} & \multicolumn{1}{c|}{R2} & \multicolumn{1}{c}{Runtime} & \multicolumn{1}{c}{RTR} & \multicolumn{1}{c}{MAE} & \multicolumn{1}{c}{R2} \\
\midrule
\texttt{aes\_core} & 22938 & 23199 & 66221 & 738.70  & 2.26  & 0.00  & 1.000  & 1092.93  & 3.34  & 0.18  & 0.995  & 326.91  & 1.00  & 0.24  & 0.988  \\
\texttt{b19} & 255278 & 255300 & 776320 & 7165.86  & 5.69  & 0.00  & 1.000  & 11246.62  & 8.92  & 10.02  & 0.977  & 1260.33  & 1.00  & 4.84  & 0.985  \\
\texttt{des\_perf} & 138878 & 139112 & 371587 & 2592.86  & 3.27  & 0.00  & 1.000  & 4558.99  & 5.75  & 0.65  & 0.981  & 792.54  & 1.00  & 0.54  & 0.993  \\
\texttt{edit\_dist} & 147650 & 150212 & 416609 & 3696.77  & 4.62  & 0.00  & 1.000  & 10136.41  & 12.66  & 1.05  & 0.978  & 800.55  & 1.00  & 1.05  & 0.984  \\
\texttt{fft} & 38158 & 39184 & 116139 & 1065.16  & 2.88  & 0.00  & 1.000  & 2049.21  & 5.54  & 0.68  & 0.979  & 370.17  & 1.00  & 0.91  & 0.973  \\
\texttt{leon2} & 1616369 & 1616984 & 4178874 & 41594.73  & 7.56  & 0.00  & 1.000  & 79924.12  & 14.53  & 6.01  & 0.963  & 5500.07  & 1.00  & 1.76  & 0.993  \\
\texttt{leon3mp} & 1247725 & 1247979 & 3267993 & 32634.40  & 7.34  & 0.00  & 1.000  & 76333.45  & 17.17  & 10.00  & 0.976  & 4446.62  & 1.00  & 4.77  & 0.985  \\
\texttt{matrix\_mult} & 164040 & 167242 & 475186 & 3124.04  & 3.26  & 0.00  & 1.000  & 9964.24  & 10.41  & 0.90  & 0.983  & 957.29  & 1.00  & 1.04  & 0.976  \\
\texttt{mgc\_edit\_dist} & 161692 & 164254 & 444693 & 5244.76  & 5.72  & 0.00  & 1.000  & 11713.04  & 12.77  & 3.58  & 0.921  & 917.38  & 1.00  & 2.40  & 0.988  \\
\texttt{mgc\_matrix\_mult} & 171282 & 174484 & 489670 & 5299.49  & 6.25  & 0.00  & 1.000  & 9520.60  & 11.22  & 2.70  & 0.932  & 848.20  & 1.00  & 2.40  & 0.969  \\
\texttt{netcard} & 1496719 & 1498555 & 3901343 & 34789.72  & 6.35  & 0.00  & 1.000  & 80598.09  & 14.72  & 9.00  & 0.984  & 5476.07  & 1.00  & 5.78  & 0.985  \\
\texttt{pci\_bridge32} & 40790 & 40950 & 108172 & 1049.68  & 2.78  & 0.00  & 1.000  & 1539.64  & 4.08  & 1.34  & 0.992  & 377.05  & 1.00  & 0.66  & 0.998  \\
\texttt{vga\_lcd} & 259067 & 259152 & 662179 & 6099.74  & 5.63  & 0.00  & 1.000  & 10167.42  & 9.39  & 6.41  & 0.988  & 1083.08  & 1.00  & 4.00  & 0.984  \\
\midrule
Average &       &       &       & 11161.22  & 4.89  & 0.00  & 1.000  & 23757.29  & 10.04  & 4.04  & 0.973  & 1781.25  & 1.00  & 2.34  & 0.985  \\
\bottomrule
\end{tabular}%
}
\end{table*}

\subsection{Timing Exception Correctness}
\label{sec:result-exception}
The timing propagation step involves handling of complex timing exceptions, which is another important source of mismatch.
By instructing PrimeTime to write out its delay calculation results, we are able to isolate the delay calculation and timing propagation steps and test them separately -- the latter of which is exactly the problem formulation of INSTA~\cite{instadac25}, our baseline.
INSTA expects a set of \texttt{.csv} files containing circuit startpoint clock periods, endpoint required arrival times, arc delays, and a set of ``timing-disabled'' pins, which we all generate using PrimeTime Tcl scripts.
The timing-disabled pin list captures some of the false path settings but not all of them, because false path exceptions may contain multiple \texttt{-through} patterns that eliminate only paths that go through a predefined pin sequence. Such patterns cannot be replaced by disabling certain pins in the graph completely.
Moreover, there are other types of exceptions unhandled in INSTA, including multi-cycle paths, set min/max delays, etc.

By generating a \texttt{.sdf} file containing all delay annotations, we are able to compare our endpoint slacks with INSTA directly.
The results are shown in Table~\ref{tab:except}.
We use two sets of design constraints for every benchmark.
The simple constraints contain only clock definitions (\texttt{create\_clock}) and port annotations (\texttt{set\_input\_delay} and \texttt{set\_output\_delay}).
The complex constraints additionally contain randomly-sampled \texttt{set\_false\_path} and \texttt{set\_multicycle\_path} exceptions, in approximately 1:5 ratio.
With the simple constraints, both INSTA and HeteroSTA can achieve 0.999 correlation with PrimeTime endpoint slacks.
However, the correlation of INSTA degrades severely given the complex constraints, down to 0.698 $R^2$, whereas HeteroSTA maintains 0.999 correlation even with timing exceptions.

\begin{table*}[t]
\caption{Comparison of endpoint slack correlation ($R^2$) between INSTA~\cite{instadac25} and HeteroSTA under simple and complex SDC exceptions. PrimeTime is set as the golden result.}
\label{tab:except}
\resizebox{\linewidth}{!}{%
\begin{tabular}{l|rrr||rrrrr||rrrrr}
\toprule
\multirow{2}[2]{*}{Benchmark} & \multicolumn{3}{c||}{Statistics} & \multicolumn{5}{c||}{Simple SDC Test} & \multicolumn{5}{c}{Complex SDC Test} \\
      & \multicolumn{1}{l}{\#Pins} & \multicolumn{1}{l}{\#Edges} & \multicolumn{1}{l||}{\#Cells} & \multicolumn{1}{l}{\#FPs} & \multicolumn{1}{l}{\#MCPs} & \multicolumn{1}{l}{PT} & \multicolumn{1}{l}{INSTA} & \multicolumn{1}{l||}{HeteroSTA} & \multicolumn{1}{l}{\#FPs} & \multicolumn{1}{l}{\#MCPs} & \multicolumn{1}{l}{PT} & \multicolumn{1}{l}{INSTA} & \multicolumn{1}{l}{HeteroSTA} \\
\midrule
tau2015\_crc32d16N & 738   & 848   & 246   & 0     & 0     & 1.000  & 0.996  & 0.998  & 0     & 2     & 1.000  & \textcolor[rgb]{ 1,  0,  0}{0.923} & 0.998  \\
usb\_phy\_ispd & 1759  & 2105  & 604   & 0     & 0     & 1.000  & 0.999  & 1.000  & 0     & 3     & 1.000  & \textcolor[rgb]{ .929,  .49,  .192}{0.962} & 1.000  \\
aes\_core & 62142 & 80420 & 21347 & 0     & 0     & 1.000  & 0.999  & 1.000  & 17    & 65    & 1.000  & 0.997  & 1.000  \\
tau2015\_softusb\_navre & 14377 & 18538 & 4653  & 0     & 0     & 1.000  & 1.000  & 1.000  & 5     & 14    & 1.000  & \textcolor[rgb]{ .929,  .49,  .192}{0.987} & 1.000  \\
pci\_bridge32\_ispd & 87435 & 106840 & 30673 & 0     & 0     & 1.000  & 0.998  & 0.999  & 180   & 911   & 1.000  & \textcolor[rgb]{ .929,  .49,  .192}{0.989} & 0.999  \\
cordic\_ispd & 120378 & 154979 & 41601 & 0     & 0     & 1.000  & 1.000  & 1.000  & 13    & 73    & 1.000  & \textcolor[rgb]{ .929,  .49,  .192}{0.953} & 1.000  \\
fft\_ispd & 101375 & 127510 & 32281 & 0     & 0     & 1.000  & 1.000  & 1.000  & 23    & 121   & 1.000  & 1.000  & 1.000  \\
tau2015\_tip\_master & 56151 & 66792 & 18851 & 0     & 0     & 1.000  & 1.000  & 1.000  & 14    & 81    & 1.000  & \textcolor[rgb]{ 1,  0,  0}{0.698} & 1.000  \\
\bottomrule
\end{tabular}%
}
  \begin{list}{}{%
    \setlength{\leftmargin}{10pt}
    \setlength{\itemsep}{2pt}
  }
  \item PT: PrimeTime. \qquad \#FPs: number of false path exceptions. \qquad \#MCPs: number of multi-cycle paths.
  \item INSTA and HeteroSTA both read in SDF delay annotations generated by PrimeTime to bypass delay calculation and only focus on timing propagation and exception handling correctness. $R^2$ scores under 0.99 but above 0.95 are colored \textcolor[rgb]{ .929,  .49,  .192}{brown}, while scores under 0.95 are colored \textcolor[rgb]{ 1,  0,  0}{red}.
  \item We note that INSTA always calculate top-256 arrival times for every pin due to its CPPR approximation trick and this feature is currently partly hard-coded into its binary CPython extension and thus we were unable to switch it off.
As a result, we can only use a set of small benchmarks from TAU 2015 contest to avoid GPU memory overflow in INSTA.
We did not report runtime because (1) the benchmarks are too small to provide valuable runtime comparison, and (2) it would be unfair for INSTA since it computes 256$\times$ more arrival times.
  \end{list}
\end{table*}

\subsection{Case Study: Timing-Driven Global Placement}
\label{sec:result-gp}
We integrate HeteroSTA into two state-of-the-art open-source timing-driven global placement flows, DREAMPlace 4.0~\cite{liao2023dreamplace} and Efficient-TDP~\cite{shi2025timing}.
Both of the flows are based on the GPU-accelerated DREAMPlace for global placement and the CPU-based OpenTimer for timing analysis.
Their flows suffer from severe runtime overhead due to frequent CPU-GPU data transfer and the translation between flattened and OOP-style data structures.

By simply substituting OpenTimer with HeteroSTA in their flows and leave other algorithms unchanged, we achieve significant end-to-end runtime speedup in both of the flows, as shown in Table~\ref{tab:gp} on ICCAD 2015 contest benchmarks~\cite{ICCAD15TimingContest}.
In DREAMPlace 4.0, the runtime of the entire flow has been accelerated by 4.5$\times$ on average.
In Efficient-TDP, the end-to-end flow runtime has been accelerated by 5.77$\times$ on average.
All these runtime speedups have come with equal or slightly better WNS and TNS at the end of their optimization loops.
These results have proven the necessity and effectiveness of HeteroSTA as a heterogeneous STA library in future fully-heterogeneous physical design flows.

We note that this also demonstrates the generality of HeteroSTA, as DREAMPlace 4.0 and Efficient-TDP use the STA engine differently.
DREAMPlace 4.0 is based on back-propagated pin slacks whereas Efficient-TDP is based on top-$k$ critical paths.
HeteroSTA supports both STA report styles efficiently and can be applied to a large range of applications.
We have released our code changes to DREAMPlace 4.0 and Efficient-TDP on GitHub as HeteroSTA integration demos and for the ease of reproducibility.

\begin{table*}[t]
\caption{Comparison of TNS ($\times 10^5$ ps), WNS ($\times 10^3$ ps), HPWL ($\times 10^6$), and runtime among DREAMPlace 4.0 \cite{liao2023dreamplace}, Efficient-TDP \cite{shi2025timing}, and their respective integrations with HeteroSTA. Runtimes are in seconds.}
\label{tab:gp}%
\resizebox{\linewidth}{!}{%
    \begin{tabular}{l|cccc|cccc|cccc|cccc}
    \toprule
     \multirow{2}{*}{\centering   Benchmark} &
     \multicolumn{4}{c|}{DREAMPlace 4.0 \space \cite{liao2023dreamplace}} & \multicolumn{4}{c|}{DREAMPlace 4.0
     \cite{liao2023dreamplace} + HeteroSTA} & 
     \multicolumn{4}{c|}{Efficient-TDP \space
     \cite{shi2025timing}} & \multicolumn{4}{c}{Efficient-TDP \cite{shi2025timing} + HeteroSTA} \\

    & TNS & WNS & HPWL & Runtime & TNS & WNS & HPWL & Runtime & TNS & WNS & HPWL & Runtime & TNS & WNS & HPWL & Runtime \\  
    \midrule
    \texttt{superblue1} & -87.91 & -14.23 & 493.86 & 526.45 & -59.62 & -12.58 & 449.19 & 112.35 & -15.71 & -7.77 & 418.73 & 594.72 & -16.39 & -7.65 & 418.83 & 129.89 \\
    \texttt{superblue3} & -48.74 & -15.10 & 483.06 & 626.18 & -59.28 & -15.52 & 476.63 & 110.15 & -19.98 & -11.72 & 462.68 & 607.01 & -19.88 & -11.96 & 462.54 & 143.63 \\
    \texttt{superblue4} & -145.90 & -12.84 & 334.05 & 231.97 & -150.46 & -12.80 & 333.91 & 75.42 & -87.09 & -9.38 & 317.47 & 1589.17 & -91.69 & -8.78 & 317.72 & 104.20 \\
    \texttt{superblue5} & -95.79 & -29.55 & 536.72 & 539.41 & -92.16 & -25.79 & 523.97 & 139.39 & -61.11 & -24.65 & 477.40 & 797.55 & -62.65 & -23.99 & 483.98 & 159.44 \\
    \texttt{superblue7} & -59.74 & -15.22 & 603.61 & 736.89 & -61.86 & -15.22 & 604.46 & 174.66 & -50.91 & -15.22 & 597.33 & 791.24 & -37.80 & -15.22 & 597.78 & 200.10 \\
    \texttt{superblue10} & -655.36 & -23.11 & 1087.73 & 917.77 & -628.81 & -22.17 & 1042.94 & 244.52 & -559.76 & -24.10 & 911.92 & 1305.33 & -567.71 & -23.60 & 911.38 & 272.70 \\
    \texttt{superblue16} & -63.69 & -10.02 & 459.08 & 318.42 & -51.93 & -11.87 & 467.60 & 55.66 & -21.91 & -9.03 & 474.88 & 300.53 & -18.51 & -9.85 & 463.77 & 81.25 \\
   \texttt{superblue18} & -46.75 & -11.53 & 248.83 & 279.61 & -47.34 & -11.73 & 244.39 & 48.80 & -16.58 & -6.73 & 234.92 & 277.85 & -15.70 & -6.83 & 234.41 & 59.36 \\
    \midrule
    Average Ratio & 2.511 & 1.313 & 1.080 & 3.712 & 2.287 & 1.292 & 1.055 & 0.813 & 1.058 & 1.002 & 1.001 & 5.772 & 1.000 & 1.000 & 1.000 & 1.000 \\
    \bottomrule
    \end{tabular}
}
  \begin{list}{}{%
    \setlength{\leftmargin}{10pt}
    \setlength{\itemsep}{2pt}
  }
  \item Our HeteroSTA integration code for both baselines are released on GitHub: https://github.com/limbo018/DREAMPlace, https://github.com/PKU-IDEA/Efficient-TDP-HeteroSTA. Efficient-TDP had a few bugs that made it non-deterministic, which we fixed to ensure reproducibility. 
  \end{list}
\end{table*}

\subsection{Case Study: Timing-Driven Global Routing}
\label{sec:result-gr}
In addition to global placement, we have also integrated HeteroSTA to other physical design flows such as GPU-accelerated global routing.
Our timing-driven router is based on the GPU-accelerated global router HeLEM-GR~\cite{zhao2024helemgr} and powered by HeteroSTA.
It has won the first place at the ISPD 2025 performance-driven large-scale global routing contest~\cite{ISPD2025Contest}.
Table~\ref{tab:gr} shows an ablation study of timing-driven optimization in the heterogeneous global routing flow.
All metrics including WNS, TNS, power, and routing congestion have been improved remarkbly with HeteroSTA in the loop providing real-time timing feedback.
Although the one-time initialization (reading netlist and constraints) takes a lot of additional time compared to the flow without timing-driven optimization, we note that optimizations such as persistent database across design stages are possible in a more realistic flow beyond the contest benchmarks.

\begin{table*}[t]
  \caption{The ablation study of timing, power, and congestion in a HeteroSTA-powered timing-driven global routing flow based on HeLEM-GR~\cite{zhao2024helemgr}.}
  \label{tab:gr}
  \resizebox{\linewidth}{!}{%
\begin{tabular}{rrr|rr|rr|rr|rr|rr|rr}
\toprule
\multicolumn{3}{c|}{Benchmark} & \multicolumn{2}{c|}{WNS} & \multicolumn{2}{c|}{TNS} & \multicolumn{2}{c|}{Power} & \multicolumn{2}{c|}{Congestion} & \multicolumn{2}{c|}{IO RT (s)} & \multicolumn{2}{c}{Optim. RT (s)} \\
\multicolumn{1}{c}{ID} & \multicolumn{1}{l}{Name} & \multicolumn{1}{l|}{\#Nets} & \multicolumn{1}{c}{w/o} & \multicolumn{1}{c|}{w} & \multicolumn{1}{c}{w/o} & \multicolumn{1}{c|}{w} & \multicolumn{1}{c}{w/o} & \multicolumn{1}{c|}{w} & \multicolumn{1}{c}{w/o} & \multicolumn{1}{c|}{w} & \multicolumn{1}{c}{Read .v} & \multicolumn{1}{c|}{Read .sdc} & \multicolumn{1}{c}{w/o} & \multicolumn{1}{c}{w} \\
\midrule
\multicolumn{1}{c}{1} & \multicolumn{1}{l}{\texttt{ariane\_v}} & 123900 & -0.503  & {-0.365} & -1263.5  & {-997.9} & {0.646} & {0.646} & {6625796} & 6662339 & 1     & 3     & 4     & 6  \\
\multicolumn{1}{c}{2} & \multicolumn{1}{l}{\texttt{bsg\_v}} & 736883 & -0.444  & {-0.388} & -10504.2  & {-9602.9} & {3.052} & 3.053  & 24503278 & {22947622} & 9     & 8     & 10    & 19  \\
\multicolumn{1}{c}{3} & \multicolumn{1}{l}{\texttt{nvdla\_v}} & 199481 & -67.563  & {-56.273} & -579481.9  & {-515380.2} & 2.942  & {2.938} & {13251396} & 13475963 & 1     & 2     & 5     & 12  \\
\multicolumn{1}{c}{4} & \multicolumn{1}{l}{\texttt{tile\_v}} & 136120 & -0.670  & {-0.387} & -3405.3  & {-1711.6} & 0.145  & {0.144} & {2260103} & 2689765 & 1     & 2     & 4     & 7  \\
\multicolumn{1}{c}{5} & \multicolumn{1}{l}{\texttt{group\_v}} & 3274611 & -0.443  & {-0.327} & -28230.2  & {-17619.7} & 7.643  & {7.582} & 70923835 & {67934255} & 31    & 32    & 44    & 79  \\
\multicolumn{1}{c}{6} & \multicolumn{1}{l}{\texttt{cluster\_v}} & 12047279 & -0.416  & {-0.247} & -85758.7  & {-52800.0} & 23.398  & {23.150} & 349322586 & {244407995} & 136   & 131   & 124   & 261  \\
\multicolumn{1}{c}{7} & \multicolumn{1}{l}{\texttt{ariane\_b}} & 105924 & -1.418  & {-0.645} & -444.0  & {-111.5} & {0.156} & {0.156} & 4473420 & {4371048} & 1     & 2     & 5     & 7  \\
\multicolumn{1}{c}{8} & \multicolumn{1}{l}{\texttt{bsg\_b}} & 768239 & {0.000} & {0.000} & {0.0} & {0.0} & {0.305} & {0.305} & 26635950 & {22506663} & 9     & 8     & 15    & 23  \\
\multicolumn{1}{c}{9} & \multicolumn{1}{l}{\texttt{nvdla\_b}} & 157744 & {0.000} & {0.000} & {0.0} & {0.0} & {0.136} & 0.136  & 13465281 & {13242875} & 1     & 1     & 6     & 9  \\
\multicolumn{1}{c}{10} & \multicolumn{1}{l}{\texttt{tile\_b}} & 135814 & -0.570  & {-0.353} & -2430.7  & {-1312.2} & {0.144} & {0.144} & 2136877 & {2087873} & 1     & 2     & 3     & 7  \\
\multicolumn{1}{c}{11} & \multicolumn{1}{l}{\texttt{group\_b}} & 3218496 & -0.661  & {-0.404} & -47936.6  & {-27116.0} & 8.357  & {8.192} & 71610825 & {68117588} & 31    & 34    & 45    & 78  \\
\multicolumn{1}{c}{12} & \multicolumn{1}{l}{\texttt{cluster\_b}} & 12168735 & -0.520  & {-0.196} & -78834.9  & {-37832.8} & 24.179  & {24.021} & 248823650 & {235469814} & 137   & 140   & 128   & 199  \\
\midrule
\multicolumn{3}{r|}{Average Ratio} & 1.000  & 0.700  & 1.000  & 0.681  & 1.000  & 0.996  & 1.000  & 0.957  &       &       & 1.000  & 1.810  \\
\bottomrule
\end{tabular}%

}
  \begin{list}{}{%
    \setlength{\leftmargin}{10pt}
    \setlength{\itemsep}{2pt}
  }
  \item w/o: without timing-driven routing; \qquad w: with HeteroSTA-powered timing-driven routing.
  \item IO RT: file I/O runtime, including netlist read and database initialization; \qquad Optim. RT: the timing-driven routing optimization loop runtime.
  \end{list}
\end{table*}


\section{Conclusion}
\label{sec:Conclusion}
This paper presents HeteroSTA, the first CPU-GPU heterogeneous STA engine with holistic industrial design support, powered by a self-contained advanced Arnoldi delay calculator, rich support for timing exceptions, and a zero-overhead heterogeneous API targeting widespread adoption in heterogeneous EDA flows.
As a STA engine, HeteroSTA provides comparable arc delay and slack correlation to leading commercial tools.
As a heterogeneous library, HeteroSTA brings remarkable end-to-end runtime speedup to various heterogeneous physical design optimization flows.
In the future, we plan to extend HeteroSTA to support other uncovered functionalities in Figure~\ref{fig:sta-in-flow}, especially CCS and other sign-off related features.

\section*{Acknowledge}
This work is supported in part by the Natural Science Foundation of Beijing, China (Grant No. Z230002), and the 111 Project (B18001).


{
\bibliographystyle{IEEEtran}
\bibliography{./ref/ieeetranbstctl,./ref/Top_sim,./ref/Timing,./ref/ALG,./ref/addition,./ref/PD,./ref/Simulation,./ref/Partition,./ref/routing,./ref/Synthesis,./ref/Software}

@article{DREAMPlace, 
    title={{DREAMPlace}: Deep Learning Toolkit-Enabled GPU Acceleration for Modern VLSI Placement},
    author={Lin, Yibo and Jiang, Zixuan and Gu, Jiaqi and Li, Wuxi and Dhar, Shounak and Ren, Haoxing and Khailany, Brucek and Pan, David Z.},
    journal   = tcad, 
    volume    = {}, 
    issue     = {}, 
    pages    = {}, 
    year      = {2020}, 
    month     = {}, 
}

@article{ABCDPlace, 
    title={{ABCDPlace}: Accelerated Batch-based Concurrent Detailed Placement on Multi-threaded CPUs and GPUs},
    author={Lin, Yibo and Li, Wuxi and Gu, Jiaqi and Ren, Haoxing and Khailany, Brucek and Pan, David Z.},
    journal   = tcad, 
    volume    = {}, 
    issue     = {}, 
    pages    = {}, 
    year      = {2020}, 
    month     = {}, 
}

@article{kahng2024dg,
  title={Dg-replace: A dataflow-driven gpu-accelerated analytical global placement framework for machine learning accelerators},
  author={Kahng, Andrew B and Wang, Zhiang},
  journal=tcad,
  year={2024},
  publisher={IEEE}
}

@article{liu2023xplace,
  title={Xplace: An extremely fast and extensible placement framework},
  author={Liu, Lixin and Fu, Bangqi and Lin, Shiju and Liu, Jinwei and Young, Evangeline FY and Wong, Martin DF},
  journal=tcad,
  volume={43},
  number={6},
  pages={1872--1885},
  year={2023},
  publisher={IEEE}
}

@inproceedings{agnesina2023autodmp,
  title={AutoDMP: Automated dreamplace-based macro placement},
  author={Agnesina, Anthony and Rajvanshi, Puranjay and Yang, Tian and Pradipta, Geraldo and Jiao, Austin and Keller, Ben and Khailany, Brucek and Ren, Haoxing},
  booktitle=ispd,
  pages={149--157},
  year={2023}
}

@inproceedings{dpo2025,
  title={Differentiable Physical Optimization},
  author={Yufan Du and Zizheng Guo and Runsheng Wang and Yibo Lin},
  booktitle=iccad,
  pages={1--6},
  year={2025},
  organization={IEEE}
}

@inproceedings{fusionsizericcad24,
 author = {Du, Yufan and Guo, Zizheng and Lin, Yibo and Wang, Runsheng and Huang, Ru},
 booktitle = iccad,
 organization = {ACM},
 title = {Fusion of Global Placement and Gate Sizing with Differentiable Optimization},
 year = {2024}
}

@inproceedings{guo2022difftdp,
author = {Guo, Zizheng and Lin, Yibo},
title = {Differentiable-timing-driven global placement},
year = {2022},
publisher = {ACM},
booktitle = dac,
pages = {1315–1320},
numpages = {6}
}

@inproceedings{sroadpgpu2025,
  title={Leveraging {GPU} for Better Detailed Placement Quality},
  author={Lu, Chen-Han and Liu, Wen-Hao and Ren, Haoxing and Wang, Ting-Chi},
  booktitle=iccad,
  pages={1--6},
  year={2025},
  organization={IEEE}
}

@article{liao2023dreamplace,
  title={{DREAMPlace} 4.0: Timing-driven placement with momentum-based net weighting and lagrangian-based refinement},
  author={Liao, Peiyu and Guo, Dawei and Guo, Zizheng and Liu, Siting and Lin, Yibo and Yu, Bei},
  journal={IEEE Transactions on Computer-Aided Design of Integrated Circuits and Systems},
  volume={42},
  number={10},
  pages={3374--3387},
  year={2023},
  publisher={IEEE}
}

@inproceedings{shi2025timing,
  title={Timing-driven global placement by efficient critical path extraction},
  author={Shi, Yunqi and Xu, Siyuan and Kai, Shixiong and Lin, Xi and Xue, Ke and Yuan, Mingxuan and Qian, Chao},
  booktitle={2025 Design, Automation \& Test in Europe Conference (DATE)},
  pages={1--7},
  year={2025},
  organization={IEEE}
}

@inproceedings{fu2024xplace_t,
    author = {Fu, Bangqi and Liu, Lixin and Wong, Martin D. F. and Young, Evangeline F. Y.},
    booktitle = iccad,
    title = {Hybrid Modeling and Weighting for Timing-driven Placement with Efficient Calibration},
    year = {2024},
}

@article{xplace_tcad,
    author={Liu, Lixin and Fu, Bangqi and Lin, Shiju and Liu, Jinwei and Young, Evangeline F.Y. and Wong, Martin D.F.},
    journal=tcad,
    title={{Xplace}: An Extremely Fast and Extensible Placement Framework},
    year={2023},
}

@inproceedings{liang2024medpart,
  title={{MedPart}: A multi-level evolutionary differentiable hypergraph partitioner},
  author={Liang, Rongjian and Agnesina, Anthony and Ren, Haoxing},
  booktitle=ispd,
  pages={3--11},
  year={2024}
}

@article{wu2025ghypart,
  title={{gHyPart}: {GPU}-friendly End-to-End Hypergraph Partitioner},
  author={Wu, Zhenlin and Zhao, Haosong and Liu, Hongyuan and Wen, Wujie and Li, Jiajia},
  journal=taco,
  volume={22},
  number={1},
  pages={1--25},
  year={2025},
  publisher={ACM}
}

@inproceedings{lee2025hyperg,
author = {Lee, Wan Luan and Lin, Dian-Lun and Chiu, Cheng-Hsiang and Schlichtmann, Ulf and Huang, Tsung-Wei},
title = {{HyperG}: Multilevel {GPU}-Accelerated k-way Hypergraph Partitioner},
year = {2025},
publisher = {ACM},
booktitle = aspdac,
pages = {1031–1040},
numpages = {10},
location = {Tokyo, Japan}
}

@INPROCEEDINGS{goodarzi2016gpupart,
  author={Goodarzi, Bahareh and Burtscher, Martin and Goswami, Dhrubajyoti},
  booktitle=ipdpsw,
  title={Parallel Graph Partitioning on a CPU-GPU Architecture},
  year={2016},
  volume={},
  number={},
  pages={58-66}}

@inproceedings{lee2024gkway,
author = {Lee, Wan Luan and Lin, Dian-Lun and Huang, Tsung-Wei and Jiang, Shui and Ho, Tsung-Yi and Lin, Yibo and Yu, Bei},
title = {G-kway: Multilevel GPU-Accelerated k-way Graph Partitioner},
year = {2024},
publisher = {ACM},
booktitle = dac,
articleno = {105},
numpages = {6},
location = {San Francisco, CA, USA}
}

@inproceedings{lee2025igkway,
  title={{iG-kway}: Incremental k-way Graph Partitioning on {GPU}},
  author={Lee, Wan Luan and Jiang, Shui and Lin, Dian-Lun and Chang, Che and Zhang, Boyang and Chung, Yi-Hua and Schlichtmann, Ulf and Ho, Tsung-Yi and Huang, Tsung-Wei},
  booktitle=dac,
  pages={1--7},
  year={2025},
  organization={IEEE}
}

@inproceedings{lin2022rtlflow,
  title={From {RTL} to {CUDA}: A {GPU} Acceleration Flow for {RTL} Simulation with Batch Stimulus},
  author={Lin, Dian-Lun and Ren, Haoxing and Zhang, Yanqing and Huang, Tsung-Wei},
  booktitle=icpp,
  year={2022}
}

@inproceedings{qian2011gpurtl,
	address = {San Jose, CA, USA},
	title = {Accelerating {RTL} simulation with {GPUs}},
	booktitle = iccad,
	publisher = {IEEE},
	author = {Qian, Hao and Deng, Yangdong},
	year = {2011},
	pages = {687--693}
}

@inproceedings{zhang2024gl0am,
	address = {New York, NY, USA},
	title = {{GL0AM}: {GPU} {Logic} {Simulation} {Using} 0-{Delay} and {Re}-simulation {Acceleration} {Method}},
  booktitle = iccad,
	language = {en},
	publisher = {IEEE},
	author = {Zhang, Yanqing and Ren, Haoxing and Khailany, Brucek},
  year={2024},
	pages = {1--9},
}

@inproceedings{gemdac25,
 author = {Guo, Zizheng and Zhang, Yanqing and Wang, Runsheng and Lin, Yibo and Ren, Haoxing},
 booktitle = dac,
 organization = {IEEE},
 title = {{GEM}: {GPU}-Accelerated Emulator-Inspired {RTL} Simulation},
 year = {2025}
}

@misc{TOOL_pt,
  title        = {Synopsys {PrimeTime}},
  howpublished = "\url{https://www.synopsys.com/implementation-and-signoff/signoff/primetime.html}",
}

@misc{TOOL_tempus,
title = {Cadence {Tempus}},
howpublished = "\url{https://www.cadence.com/en_US/home/tools/digital-design-and-signoff/silicon-signoff/tempus-timing-signoff-solution.html}",
}

@inproceedings{novelrewrite2022,
author = {Lin, Shiju and Liu, Jinwei and Liu, Tianji and Wong, Martin D. F. and Young, Evangeline F. Y.},
title = {NovelRewrite: node-level parallel AIG rewriting},
year = {2022},
publisher = {ACM},
address = {New York, NY, USA},
booktitle = dac,
pages = {427–432},
numpages = {6}
}

@inproceedings{iccadcadcontest2021c_logicrewrite,
  title={2021 {ICCAD} {CAD} contest problem {C}: {GPU} accelerated logic rewriting},
  author={Pasandi, Ghasem and Pratty, Sreedhar and Brown, David and Zhang, Yanqing and Ren, Haoxing and Khailany, Brucek},
  booktitle=iccad,
  pages={1--6},
  year={2021},
  organization={IEEE}
}

@inproceedings{sun2024massively,
  title={Massively Parallel {AIG} Resubstitution},
  author={Sun, Yang and Liu, Tianji and Wong, Martin DF and Young, Evangeline FY},
  booktitle=dac,
  pages={1--6},
  year={2024}
}

@inproceedings{liu2023rethinking,
  title={Rethinking {AIG} resynthesis in parallel},
  author={Liu, Tianji and Young, Evangeline FY},
  booktitle=dac,
  pages={1--6},
  year={2023},
  organization={IEEE}
}

@INPROCEEDINGS{finemap2024,
  author={Liu, Tianji and Chen, Lei and Li, Xing and Yuan, Mingxuan and Young, Evangeline F.Y.},
  booktitle=aspdac,
  title={{FineMap}: A Fine-grained {GPU}-parallel {LUT} Mapping Engine},
  year={2024},
  volume={},
  number={},
  pages={392-397}}

@inproceedings{groot2025,
  title={{GROOT}: Graph Edge Re-growth and Partitioning for the Verification of Large Designs in Logic Synthesis},
  author={Kiran Thorat and Hongwu Peng and Yuebo Luo and Xi Xie and Shaoyi Huang and Amit Hasan and Jiahui Zhao and Yingjie Li and Zhijie Shi and Cunxi Yu and Caiwen Ding},
  booktitle=iccad,
  pages={1--6},
  year={2025},
  organization={IEEE}
}

@misc{OpenSTA,
  title = {Open{STA}},
  howpublished = {\url{https://github.com/The-OpenROAD-Project/OpenSTA}}
}

@INPROCEEDINGS{ICCAD15TimingContest,
  author={M. {Kim} and J. {Hu} and J. {Li} and N. {Viswanathan}},
  booktitle=iccad, 
  title={{ICCAD-2015 CAD contest in incremental timing-driven placement and benchmark suite}}, 
  year={2015},
  volume={},
  number={},
  pages={921-926},}

@book{STABook,
author = {Bhasker, J. and Chadha, Rakesh},
title = {Static Timing Analysis for Nanometer Designs: A Practical Approach},
year = {2009},
isbn = {0387938192},
publisher = {Springer Publishing Company, Incorporated},
edition = {1st}
}

@article{huang2020opentimerv2,
  author={T. {Huang} and G. {Guo} and C. {Lin} and M. D. F. {Wong}},
  journal=tcad, 
  title={{OpenTimer v2: A New Parallel Incremental Timing Analysis Engine}}, 
  year={2021},
  volume={40},
  number={4},
  pages={776-786},}

@inproceedings{hu2015tau,
  title={{TAU 2015 contest on incremental timing analysis}},
  author={Hu, Jin and Schaeffer, Greg and Garg, Vibhor},
  booktitle=iccad,
  pages={882--889},
  year={2015},
  organization={IEEE}
}

@inproceedings{guo20gpusta,
 author = {Guo, Zizheng and Huang, Tsung-Wei and Lin, Yibo},
 title = {{GPU}-Accelerated Static Timing Analysis},
 booktitle = iccad,
 organization = {ACM},
 year = {2020}
}

@inproceedings{guo21heterocppr,
 author = {Guo, Zizheng and Huang, Tsung-Wei and Lin, Yibo},
 title = {{HeteroCPPR}: Accelerating Common Path Pessimism Removal with Heterogeneous {CPU-GPU} Parallelism},
 booktitle = iccad,
 organization = {ACM/IEEE},
 year = {2021}
}

@inproceedings{guo21gpupba,
 author = {Guo, Guannan and Huang, Tsung-Wei and Lin, Yibo and Wong, Martin},
 title = {{GPU}-accelerated Path-based Timing Analysis},
 booktitle = dac,
 organization = {ACM},
 year = {2021}
}

@article{gpupbatcad23,
 author = {Guo, Guannan and Huang, Tsung-Wei and Lin, Yibo and Guo, Zizheng and Yellapragada, Sushma and Wong, Martin DF},
 journal = tcad,
 publisher = {IEEE},
 title = {A {GPU}-accelerated Framework for Path-based Timing Analysis},
 year = {2023}
}

@article{gputimertcad23,
 author = {Guo, Zizheng and Huang, Tsung-Wei and Lin, Yibo},
 journal = tcad,
 number = {},
 pages = {1-1},
 title = {Accelerating Static Timing Analysis using {CPU-GPU} Heterogeneous Parallelism},
 volume = {},
 year = {2023}
}

@inproceedings{arnoldidate24,
 author = {Guo, Zizheng and Huang, Tsung-Wei and Jin, Zhou and Zhuo, Cheng and Lin, Yibo and Wang, Runsheng and Huang, Ru},
 booktitle = date,
 title = {Heterogeneous Static Timing Analysis with Advanced Delay Calculator},
 year = {2024}
}

@inproceedings{heteroexcepticcad24,
 author = {Guo, Zizheng and Zhang, Zuodong and Li, Wuxi and Huang, Tsung-Wei and Shi, Xizhe and Du, Yufan and Lin, Yibo and Wang, Runsheng and Huang, Ru},
 booktitle = iccad,
 organization = {ACM},
 title = {{HeteroExcept}: A {CPU}-{GPU} Heterogeneous Algorithm to Accelerate Exception-aware Static Timing Analysis},
 year = {2024}
}

@inproceedings{pathgenaspdac25,
 author = {Chang, Che and Zhang, Boyang and Chiu, Cheng-Hsiang and Lin, Dian-Lun and Chung, Yi-Hua and Lee, Wan-Luan and Guo, Zizheng and Lin, Yibo and Huang, Tsung-Wei},
 booktitle = aspdac,
 organization = {IEEE},
 title = {{PathGen}: An Efficient Parallel Critical Path Generation Algorithm},
 year = {2025}
}

@inproceedings{instadac25,
 author = {Lu, Yi-Chen and Guo, Zizheng and Kunal, Kishor and Liang, Rongjian and Ren, Haoxing},
 booktitle = dac,
 organization = {IEEE},
 title = {{INSTA}: An Ultra-Fast, Differentiable, Statistical Static Timing Analysis Engine for Industrial Physical Design Applications},
 year = {2025}
}

@inproceedings{lin2024gcstimer,
author = {Lin, Shiju and Guo, Guannan and Huang, Tsung-Wei and Sheng, Weihua and Young, Evangeline and Wong, Martin},
title = {{GCS}-Timer: {GPU}-Accelerated Current Source Model Based Static Timing Analysis},
year = {2024},
publisher = {ACM},
booktitle = dac,
articleno = {71},
numpages = {6}
}

@inproceedings{incregpusta2025,
  title={{IncreGPUSTA}: {GPU}-Accelerated Incremental Static Timing Analysis for Iterative Design Flows},
  author={Liu, Haichuan and Guo, Zizheng and Wang, Runsheng and Lin, Yibo},
  booktitle=iccad,
  pages={1--6},
  year={2025},
  organization={IEEE}
}

@string{aspdac  = "IEEE/ACM Asia and South Pacific Design Automation Conference (ASPDAC)"}

@string{dac     = "ACM/IEEE Design Automation Conference (DAC)"}

@string{date    = "IEEE/ACM Proceedings Design, Automation and Test in Eurpoe (DATE)"}

@string{iccad   = "IEEE/ACM International Conference on Computer-Aided Design (ICCAD)"}

@string{ispd    = "ACM International Symposium on Physical Design (ISPD)"}

@string{tau     = "ACM International Workshop on on Timing Issues in the Specification and Synthesis of Digital Systems (TAU) "}

@string{taco    = "ACM Transactions on Architecture and Code Optimization (TACO)"}

@string{tcad    = "IEEE Transactions on Computer-Aided Design of Integrated Circuits and Systems (TCAD)"}

@string{ipdpsw   = "International Parallel \& Distributed Processing Symposium Workshops (IPDPSW)"}

@string{icpp    = "International Conference on Parallel Processing (ICPP)"}

@string{open     = "Optimization and Engineering"}

@string{test     = "Test"}

@string{net      = "Networks"}

@string{aspdac   = "Proc.~ASPDAC"}

@string{dac      = "Proc.~DAC"}

@string{date     = "Proc.~DATE"}

@string{iccad    = "Proc.~ICCAD"}

@string{ispd     = "Proc.~ISPD"}

@string{tau      = "Proc.~TAU"}

@string{taco     = "ACM TACO"}

@string{tcad     = "IEEE TCAD"}

@string{ipdpsw   = "Proc.~IPDPS Workshops"}

@string{icpp    = "Proc.~ICPP"}

@inproceedings{he2022xcheck,
  title={{X-Check}: {GPU}-accelerated design rule checking via parallel sweepline algorithms},
  author={He, Zhuolun and Ma, Yuzhe and Yu, Bei},
  booktitle=iccad,
  pages={1--9},
  year={2022}
}

@inproceedings{he2023opendrc,
  title={{OpenDRC}: An efficient open-source design rule checking engine with hierarchical {GPU} acceleration},
  author={He, Zhuolun and Zuo, Yihang and Jiang, Jiaxi and Zheng, Haisheng and Ma, Yuzhe and Yu, Bei},
  booktitle=dac,
  pages={1--6},
  year={2023},
  organization={IEEE}
}

@IEEEtranBSTCTL{ieee:norepeatednames,
CTLdash_repeated_names = "no"
}

@IEEEtranBSTCTL{ieee:etal8,
CTLuse_forced_etal = "yes",
CTLmax_names_forced_etal = "8"
}

@article{lin2022gamer,
  title={{GAMER}: {GPU}-accelerated maze routing},
  author={Lin, Shiju and Liu, Jinwei and Young, Evangeline FY and Wong, Martin DF},
  journal=tcad,
  volume={42},
  number={2},
  pages={583--593},
  year={2022},
  publisher={IEEE}
}

@inproceedings{gpufluteiccad22,
 author = {Guo, Zizheng and Gu, Feng and Lin, Yibo},
 booktitle = iccad,
 organization = {ACM},
 title = {{GPU}-Accelerated Rectilinear Steiner Tree Generation},
 year = {2022}
}

@article{liu2022fastgr,
  title={{FastGR}: Global routing on {CPU--GPU} with heterogeneous task graph scheduler},
  author={Liu, Siting and Pu, Yuan and Liao, Peiyu and Wu, Hongzhong and Zhang, Rui and Chen, Zhitang and Lv, Wenlong and Lin, Yibo and Yu, Bei},
  journal=tcad,
  volume={42},
  number={7},
  pages={2317--2330},
  year={2022},
  publisher={IEEE}
}

@inproceedings{lin2024instantgr,
  title={{InstantGR}: Scalable {GPU} parallelization for global routing},
  author={Lin, Shiju and Xiao, Liang and Liu, Jinwei and Young, Evangeline FY},
  booktitle=iccad,
  pages={1--8},
  year={2024}
}

@inproceedings{li2024dgr,
  title={{DGR}: Differentiable global router},
  author={Li, Wei and Liang, Rongjian and Agnesina, Anthony and Yang, Haoyu and Ho, Chia-Tung and Rajaram, Anand and Ren, Haoxing},
  booktitle=dac,
  pages={1--6},
  year={2024}
}

@article{xiao2025instantgr,
  title={{InstantGR}: Scalable {GPU} Parallelization for 3-D Global Routing},
  author={Xiao, Liang and Lin, Shiju and Liu, Jinwei and Duan, Qinkai and Ho, Tsung-Yi and Young, Evangeline FY},
  journal=tcad,
  year={2025},
  publisher={IEEE}
}

@inproceedings{liang2025ispd,
  title={{ISPD} 2025 Performance-Driven Large Scale Global Routing Contest},
  author={Liang, Rongjian and Agnesina, Anthony and Liu, Wen-Hao and Liberty, Matt and Chang, Hsin-Tzu and Ren, Haoxing},
  booktitle=ispd,
  pages={252--256},
  year={2025}
}

@inproceedings{zhao2024helemgr,
  title={{HeLEM-GR}: Heterogeneous Global Routing with Linearized Exponential Multiplier Method},
  author={Zhao, Chunyuan and Guo, Zizheng and Wang, Rui and Wen, Zaiwen and Liang, Yun and Lin, Yibo},
  booktitle=iccad,
  pages={1--9},
  year={2024}
}

@inproceedings{zhao2025gta,
  title={{GTA}: {GPU}-Accelerated Track Assignment with Lightweight Lookup Table for Conflict Detection},
  author={Zhao, Chunyuan and Wang, Jiarui and Jiang, Xun and Lou, Jincheng and Lin, Yibo},
  booktitle=iccad,
  pages={1--9},
  year={2025}
}

@inproceedings{ISPD2025Contest,
author = {Liang, Rongjian and Agnesina, Anthony and Liu, Wen-Hao and Liberty, Matt and Chang, Hsin-Tzu and Ren, Haoxing},
title = {Invited: {ISPD} 2025 Performance-Driven Large Scale Global Routing Contest},
year = {2025},
publisher = {ACM},
address = {New York, NY, USA},
booktitle = ispd,
numpages = {5},
location = {Austin, TX, USA}
}
}

\end{document}